\documentclass[twocolumn,showpacs,aps,prd,nofootinbib,amsmath,amssymb,nobibnotes,floatfix]{revtex4-1}

\usepackage{hyperref}
\usepackage{amsfonts}
\usepackage{amsmath}
\usepackage{mathrsfs}
\usepackage{epsfig}
\usepackage{graphicx}               
\usepackage{url}
\usepackage{hyperref}
\usepackage{float}
\usepackage{color}
\usepackage[utf8]{inputenc}

\usepackage{graphicx}
\usepackage{epsf}

\usepackage{comment}

\newcommand{\nc}{\newcommand}

\nc{\beq}{\begin{equation}}
\nc{\eeq}{\end{equation}}
\nc{\beqa}{\begin{eqnarray}}
\nc{\eeqa}{\end{eqnarray}}

\usepackage{slashed}

\newcommand{\lsim}{\!\mathrel{\hbox{\rlap{\lower.55ex \hbox{$\sim$}} \kern-.34em \raise.4ex \hbox{$<$}}}}
\newcommand{\gsim}{\!\mathrel{\hbox{\rlap{\lower.55ex \hbox{$\sim$}} \kern-.34em \raise.4ex \hbox{$>$}}}}

\def\be{\begin{equation}}
\def\ee{\end{equation}}

\newcommand\affspc{\vspace{4pt}}

\usepackage{footmisc}
\usepackage{setspace}

\begin{document}

\title{Superradiant instability of massive vector fields around spinning black holes in the
relativistic regime}

\author{William E.\ East}
\affiliation{Perimeter Institute for Theoretical Physics, Waterloo, Ontario N2L 2Y5, Canada \affspc}

\begin{abstract}
We study the superradiant instability of massive vector fields, i.e. Proca
fields, around spinning black holes in the test field limit.  This is motivated
by the possibility that observations of astrophysical black holes can probe the
existence of ultralight bosons subject to this mechanism.  By making use of
time-domain simulations, we characterize the growth rate, frequency, spatial
distribution, and other properties of the unstable modes, including in the
regime where the black hole is rapidly spinning and the Compton wavelength of
the Proca field is comparable to the black hole radius.  We find that
relativistic effects in this regime increase the range of Proca masses that are
unstable, as well as the maximum instability rate.  We also study the
gravitational waves that can be sourced by such an instability, finding that
they can be significantly stronger than in the massive scalar field case. 
\end{abstract}

\maketitle

\section{Introduction}
An intriguing possibility, that has been the focus of recent interest, is that
spinning black holes could, in fact, be unstable because of the existence of
ultralight massive bosons. This so-called superradiant instability relies on
the wave analogue of the Penrose process, whereby waves with frequency $\omega<m
\Omega_H$, where $m$ is the azimuthal number of the wave and $\Omega_H$ is the
black hole horizon frequency, can superradiantly scatter off the black hole,
gaining energy and angular momentum in the process.  When this is combined with
the fact that massive bosons can form gravitationally bound states around black
holes, the ``black hole bomb" originally proposed
in~\cite{1972Natur.238..211P} is realized: small (even vacuum) fluctuations of a massive boson can be
exponentially amplified into large clouds that potentially gain some
non-negligible fraction of the black hole's rotational
energy~\cite{Damour:1976,Detweiler:1980uk,Zouros:1979iw}.  

Hence, the superradiant instability allows astrophysical black holes to act as
particle detectors of sorts, providing a unique method to look for
weakly coupled, ultralight massive bosons such as the string
axiverse~\cite{Arvanitaki:2009fg,Arvanitaki:2010sy}, the QCD
axion~\cite{Arvanitaki:2014wva,Arvanitaki:2016qwi}, or dark/massive
photons~\cite{Holdom:1985ag,Pani:2012bp}.  
The superradiant instability is strongest for particles with
a Compton wavelength that is comparable to the radius of a given black hole,
where particles masses in the range $\sim10^{-10}$ to $10^{-18}$ eV are probed
by black holes with masses $\sim1$ to $10^8$ $M_{\odot}$.  
In light of the recent observations of gravitational waves (GWs) from merging black 
holes by LIGO~\cite{Abbott:2016blz,Abbott:2016nmj}, with many more expected in the
coming years as LIGO reaches design sensitivity~\cite{Abbott:2016nhf},
measuring black hole masses and spins from GW observations
is one promising avenue for ruling out, or providing evidence for, the existence
of such ultralight massive fields~\cite{Arvanitaki:2016qwi,BLT2017}; 
in addition the oscillating bosonic clouds that form
from the superradiant instability can be a source of nearly monochromatic
GWs, which could also be detectable.  

There have been numerous studies of the superradiant instability of massive
scalar fields around spinning black
holes~\cite{Dolan:2007mj,Brito:2014wla,Arvanitaki:2010sy,Arvanitaki:2014wva,Arvanitaki:2016qwi},
including time-domain simulations~\cite{Dolan:2012yt,Witek:2012tr}, and studies
of the gravitational radiation produced~\cite{Yoshino:2013ofa}.  Though massive
vector fields, i.e. Proca fields, can have significantly faster rates of
superradiant growth around a spinning black hole than scalar fields, their
treatment is more difficult since, even at the level of a test field, the
equations do not decouple in the Teukolsky formalism.  There have been studies
specializing to a nonspinning black hole~\cite{Galtsov:1984ixy,Konoplya:2005hr,Konoplya:2006gq,Rosa:2011my},
making use of a slow rotation approximation for the black
hole~\cite{Pani:2012bp,Pani:2012vp}, using effective field theory
methods~\cite{Endlich:2016jgc}, and recent work using a matching
procedure~\cite{BLT2017} to analytically calculate the instability rate in the
nonrelativistic limit.  However, a complete picture of the superradiant
instability for massive vectors in the relativistic regime, where the Compton
wavelength of the field is comparable to the black hole radius---which is
also the regime where the instability rate will be largest---has been
lacking.

Here we attempt to address this by making use of time-domain simulations of the
full Proca equations on Kerr spacetimes.  In~\cite{Witek:2012tr}, a full $3+1$
simulation of such a case was performed, illustrating the superradiant
instability in the relativistic regime.  (See also~\cite{Zilhao:2015tya} for a
study of the nonlinear behavior of Proca fields around nonspinning black
holes.) However, the computational expense of such simulations makes it
difficult to follow such evolutions for very long times, or to explore the
parameter space.  Here we use a similar approach, but by fixing the azimuthal
symmetry of the field, we are able to reduce the computational domain to two
dimensions, which allows us to evolve the fields for time scales of $\sim10^5$
times the horizon crossing time of the black hole, and thus to clearly
determine the properties of the superradiant instability across a range of
parameters.  We find that the Proca field superradiant instability growth rate
can be quite fast, with, for example, the Proca field energy having a minimum
$e$-folding time of $\sim70 (M/10 M_{\odot})$ ms for a black hole spin of mass $M$
and dimensionless spin $a=0.99$ (for a Proca field with azimuthal number
$m=1$), and about 20 longer for the next lowest azimuthal number ($m=2$).  We
show that relativistic effects extend the range of mass parameters over which
the Proca field is unstable by decreasing the frequency of the bound states
relative to the nonrelativistic approximation, and come up with a simple
approximation for these corrections.  We also study the properties of the
superradiantly unstable Proca field modes, including the gravitational
radiation that they can source, finding that the GW luminosity in the
relativistic regime can be much higher than in the equivalent scalar field
case.

The rest of the paper is organized as follows.  In Sec.~\ref{methods} we
summarize the Proca field equations and the numerical methods we use for
evolving them, and define the quantities used for analyzing the superradiant
instability; in Sec.~\ref{results} we present results for the superradiant
instability including the growth rate, frequency, mass range frequency, spatial
distribution, and GW signal of the unstable modes; in Sec.~\ref{discussion} we
discuss these results, compare them to others in the literature, and conclude
with some directions for future work; in the Appendix we provide some details
on numerical convergence and error estimates.

\section{Methods}
\label{methods}
In this paper, we study the evolution of Proca fields on a fixed Kerr spacetime
with mass $M$ and dimensionless spin $a$.  We use Cartesian Kerr-Schild
coordinates~\cite{1965cngg.conf..222K} where the spacetime metric is given by
\beqa
ds^2 &=&-dt^2+dx^2+dy^2+dz^2+\nonumber \\
&&\frac{2Mr^3}{r^4+a^2M^2z^2}\big[
dt+\frac{z}{r}dz
+\frac{r}{r^2+a^2M^2}(xdx+ydy) \nonumber \\ &&-\frac{aM}{r^2+a^2M^2}(xdy-ydx)\big]^2, 
\nonumber
\eeqa
where $(x^2+y^2)/(r^2+a^2)+z^2/r^2=1$.  Here and throughout we use units with
$G=c=1$.

We consider a complex Proca field $X_a=X_a^R+iX_a^I$ (or equivalently, two
uncoupled real Proca fields) with equation of motion
\beq
\nabla_a F^{ab}= \mu^2 X^b,
\eeq
where $F_{ab}=\nabla_a X_b-\nabla_b X_a$ and $\mu$ is the Proca field mass.  

Following~\cite{Choptuik:2003ac}, in order to make the problem numerically
tractable, we will assume that the Proca field has a specific azimuthal
dependence which will allow us to make our computational domain two
dimensional. Let $\mathcal{L}_{\phi}$ be the Lie derivative with respect to the
usual axisymmetric Killing vector of the Kerr spacetime.  Then we will assume
that 
\beq
\mathcal{L}_{\phi}X_a=imX_a, 
\label{eqn:asym}
\eeq
where $m$ is a non-negative integer (below we will study $m=1$ and 2).  We then
use a generalization of the modified cartoon method introduced
in~\cite{Pretorius:2004jg} where we take our numerical domain to be the
two-dimensional half plane given by $0\leq x < \infty$ and $-\infty < z <
\infty$.  Derivatives in the $y$ direction are calculated by rewriting them in
terms of $x$ and $z$ derivatives using Eq.~\ref{eqn:asym}, and regularity is
imposed at the $z$ axis.
In Sec.~\ref{ssec:gw} we also perform a few shorter simulations of the Einstein-Proca
equations with just a real-valued Proca field in a fully three-dimensional computational domain
to study GW production.

As in~\cite{Zilhao:2015tya}, we use a 3+1 decomposition of $X_a$ into a scalar
$\chi$ and a three vector $\chi_{i}$,
\beq
X_a = \chi_a+n_a\chi,
\eeq
where $n_a$ is the unit vector perpendicular to slices of constant coordinate
time, and introduce an electric field
\beq
E_i = \gamma^{a}_i F_{ab} n^b,
\eeq
where $\gamma^a_b=\delta^a_b+n^an_b$ is the spatial projection operator.  Also
as in~\cite{Zilhao:2015tya}, we introduce an auxiliary scalar field $Z$ to dampen
violations of the Proca field-equivalent of the Gauss constraint at some
prescribed rate $\sigma$.  We then evolve the real and imaginary parts of the
variables $\{\chi,\chi_i,E^i,Z\}$ according to
\beqa
\alpha^{-1}(\partial_t - \mathcal{L}_{\beta}) \chi_i &=& -E_i-\partial_i \chi -\chi \partial_i \log\alpha,\\
\alpha^{-1}(\partial_t - \mathcal{L}_{\beta}) \chi &=& K\chi -D_i \chi^i -\chi^i \partial_i \log\alpha -Z,\\
\alpha^{-1}(\partial_t - \mathcal{L}_{\beta}) E^i &=& K E^i +D^iZ+\mu^2\chi^i +\nonumber \\ && \epsilon^{ijk}D_jB_k - \epsilon^{ijk}B_j\partial_k \log\alpha,\\
\alpha^{-1}(\partial_t - \mathcal{L}_{\beta}) Z &=& D_i E^i +\mu^2 \chi -\sigma Z,
\eeqa
where $\alpha$ and $\beta^i$ are the lapse and shift, respectively, $K$ is the trace of the
extrinsic curvature, $D_i$ is the covariant derivative associated with the
spatial metric, $\epsilon_{ijk}$ is the spatial totally antisymmetric
tensor, and $B^i=\epsilon^{ijk}D_j \chi_k$ is the magnetic field.  
We evolve these equations on the Kerr spacetime using fourth-order
Runge-Kutta time stepping and standard fourth-order stencils for spatial
derivatives.

For our numerical grid, we use seven levels of mesh refinement, with 2:1
refinement ratio, centered on the black hole.  For the results presented below,
we use a grid with $96\times192$ points on each mesh refinement level and a
corresponding resolution of $dx/M\approx 0.0256$ on the finest level, or
better.  More details on the resolution, as well as numerical convergence and
error estimates are given in the Appendix.

\subsection{Measured quantities}
The energy-momentum tensor associated with the Proca fields is given
by 
\beqa
T_{ab}&=&\frac{1}{2}(F_{ac}\bar{F}_{bd}+\bar{F}_{ac}F_{bd})g^{cd}-\frac{1}{4}g_{ab}F_{cd}\bar{F}^{cd} \nonumber \\
&&+\frac{\mu^2}{2}(X_a\bar{X}_b+\bar{X}_aX_b-g_{ab}X_c\bar{X}^c),
\eeqa
where the overbar indicates complex conjugation.  Using the time 
and axisymmetric Killing vectors of the Kerr spacetime, we can define
an energy
\beq
E := \int -T^t_t\sqrt{-g} d^3x := \int \rho_E \sqrt{\gamma} d^3x
\eeq 
and angular momentum 
\beq
J := \int T^t_\phi\sqrt{-g} d^3x := \int \rho_J \sqrt{\gamma} d^3x
\eeq 
for the Proca field (where $g$ and $\gamma$ are the determinants of the four- and 
three-metric, respectively), along with associated energy and angular momentum
densities on slices of constant coordinate time.  We will make use of these quantities,
evaluated in the exterior of the black hole horizon.  Any change
in $E$ and $J$ will be due to a flux of these quantities through the black hole horizon,
which can be calculated in terms of surface integrals over the horizon:
\beq
\dot{E}^H = \int -\alpha T^i_t dA_i  
\eeq
and
\beq
\dot{J}^H = \int \alpha T^i_\phi dA_i .  
\eeq

\section{Results}
\label{results}

\subsection{Onset of linear instability} 
\begin{figure}
\begin{center}
\includegraphics[width=\columnwidth,draft=false]{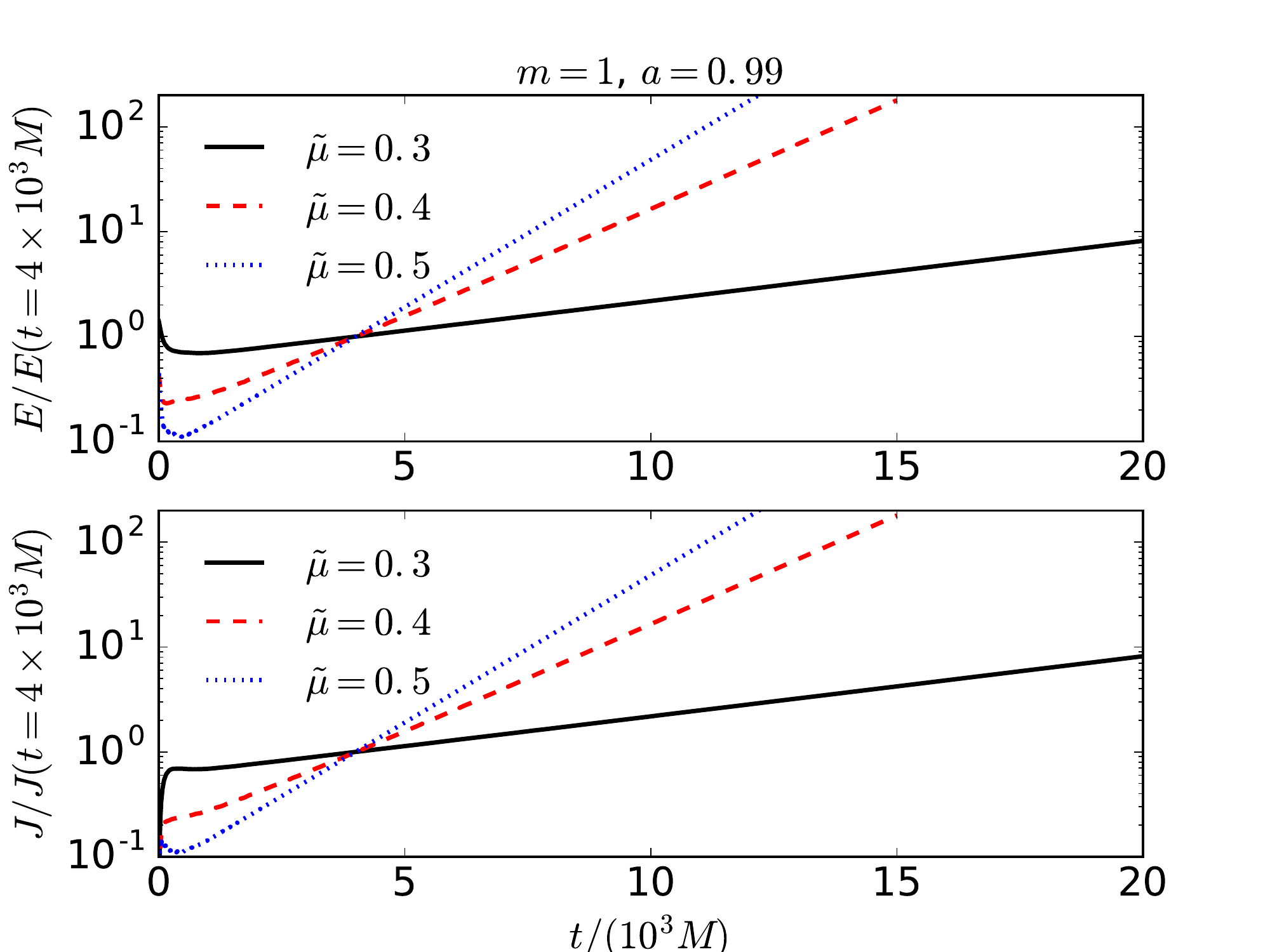}
\end{center}
\caption{
The energy (top) and angular momentum (bottom) in the Proca field as a function
of time for cases with $m=1$ and a black hole with spin $a=0.99$.  After a short transient
period, the evolution is dominated by an exponentially growing mode.
\label{fig:linear_growth}
}
\end{figure}

\begin{figure}
\begin{center}
\includegraphics[width=\columnwidth,draft=false]{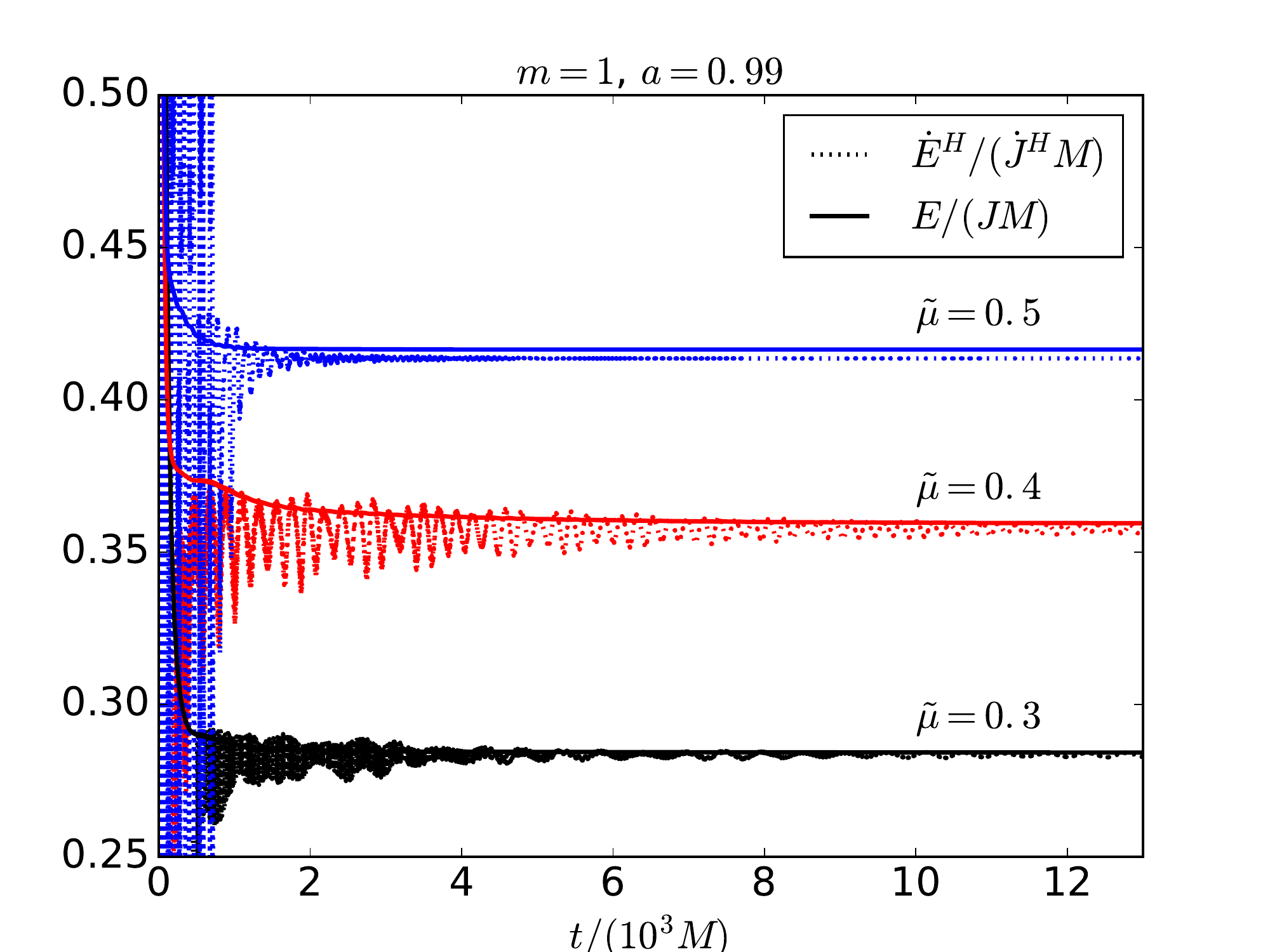}
\end{center}
\caption{
The ratio of the energy to angular momentum in the Proca field $E/J$, 
and the ratio of the energy flux to angular momentum flux
$\dot{E}^H/\dot{J}^H$ through the black horizon, as a function of
time for cases with $\tilde{\mu}=0.3$, 0.4, and 0.5, $m=1$, and a black hole with spin $a=0.99$.
After the onset of the instability, this ratio settles to a roughly constant value
which can be used as a measure of the frequency $\omega_R$ of the dominant unstable mode.
\label{fig:ej_ratio}
}
\end{figure}

In order to study the superradiant instability we start with some, essentially
arbitrary, initial data and evolve it until the solution becomes dominated by
the most unstable mode (consistent with the specified azimuthal symmetry).  We
illustrate this in Fig.~\ref{fig:linear_growth} where we show the energy and
angular momentum of the Proca field as a function of time for several cases
with $a=0.99$, $m=1$, and $\tilde{\mu}:=M\mu=0.3$, 0.4, and 0.5.  After a
transient phase, the solution clearly settles into exponential growth allowing
for the instability growth rate and other properties of the dominant unstable
mode to be measured.  For these cases we find that the energy and angular
momentum grow as $e^{2 \omega_I t}$ (where the factor of 2 is included since
the energy and angular momentum are quadratic in the fields) with
$M\omega_I\approx 7\times10^{-5}$, $2\times10^{-4}$, and $3\times10^{-4}$ for
$\tilde{\mu}=0.3$, 0.4, and 0.5, respectively.  Here we use initial data given
by $\chi_x=-i\chi_y=Ae^{-r/r_0}/\gamma^{1/2}$ and $\chi_z=\chi=E^i=0$, though
other configurations give similar results (but may take longer for the most
unstable mode to clearly emerge).  As shown in Fig.~\ref{fig:ej_ratio}, as the
most unstable mode dominates, the ratio of Proca field energy to angular
momentum becomes roughly constant.  Correspondingly, the ratio of the flux in
energy and angular momentum across the black horizon settles to another
(albeit somewhat noisier) measure of this same value.  This can be used as a
measure of the real part of the frequency $\omega_R$ of the unstable mode.

In the remainder of this paper, we will concentrate on properties of the
dominant unstable mode for various values of the Proca field mass, $0.18 \leq
\tilde{\mu} \leq 1.1$; black hole spin, $0.7\leq a \leq 0.99$; and azimuthal
number, $m=1$ and 2.  For comparison, we recall the results in the literature obtained in
the limit of $\tilde{\mu} \ll 1$, where the bound modes of the Proca field
resemble that of a hydrogen
atom~\cite{Galtsov:1984ixy,Rosa:2011my,Pani:2012bp,Endlich:2016jgc,BLT2017}.
Since here we look for the fastest growing mode with a given value of $m$ (1 or
2) we expect to see the $m=j=\ell+1$ modes where $j$ and $\ell$ are,
respectively, the total and angular momentum quantum numbers (using the
notation of~\cite{BLT2017}).  These particular bounds states have real
frequency given by $\omega_R\approx\mu[1-(\tilde{\mu}/m)^2/2]$ and a growth
rate that is proportional to $\tilde{\mu}^{4m+3}M^{-1}$ for sufficiently small $\tilde{\mu}$.  The
mode functions fall off exponentially with characteristic length scale
$(M\mu^2)^{-1}$, and the dominant $m=1$ mode has spatial vector $X_i \propto
e^{-rM\mu^2}(\hat{x}+i\hat{y})e^{-i \omega t}$. 

\subsection{Frequency and growth rate}
\begin{figure}
\begin{center}
\includegraphics[width=\columnwidth,draft=false]{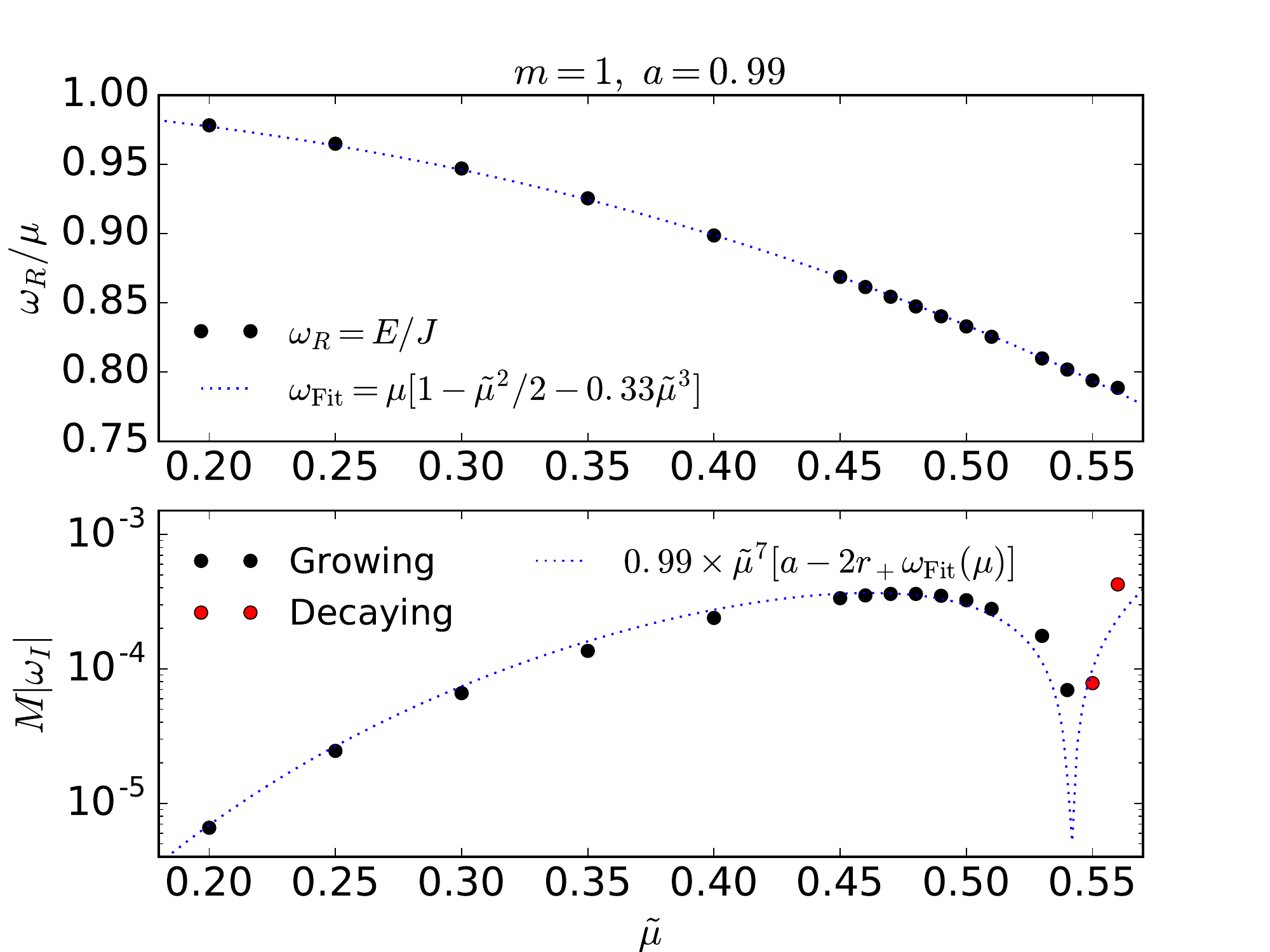}
\end{center}
\caption{
Measured values of the frequency of the dominant unstable mode for $m=1$,
$a=0.99$, and various values of $\tilde{\mu}$.  Top panel: The real part of the
frequency as measured from the ratio of the energy to angular momentum along
with the fitting function given by Eq.~(\ref{eqn:omegafit}).  Bottom panel: The
growth (black points) or decay (red points) rate of the dominant mode of the
Proca field along with the fitting function given by Eq.~(\ref{eqn:omegaI}). 
\label{fig:omega_a99}
}
\end{figure}

To begin, we focus on cases with azimuthal dependence $m=1$ and nearly extremal black hole
spin $a=0.99$---both of which maximize the growth rate of the superradiant instability---and consider a range of Proca field masses.  The results for the frequency
$\omega_R:=E/J$ and growth rate are shown in Fig.~\ref{fig:omega_a99}. For smaller values
of $\tilde{\mu}$ we find that $\omega_R/\mu\approx 1-\tilde{\mu}^2/2$ as expected in the nonrelativistic limit
where the spectrum of the bound states resembles that of a hydrogen atom in quantum
mechanics~\cite{Dolan:2007mj}. Furthermore, we find
a faster decrease in $\omega_R/\mu$ for larger $\mu$ which is well approximated by
the addition of a $\mu^3$ term, i.e. by
\beq
\omega_{\rm Fit} = \mu\left[1-(\tilde{\mu}/m)^2/2-0.33(\tilde{\mu}/m)^3\right],
\label{eqn:omegafit}
\eeq
where the last coefficient was found by fitting to the measured points. 
Here $m=1$, but we keep $m$ in the expression for use below, as we find that
this fitting form also approximates the $m=2$ data, as well as lower spins.

As shown in the bottom panel of Fig.~\ref{fig:omega_a99}, we find that Proca
fields with $\tilde{\mu}\lesssim 0.54$ are unstable, while for large values of
$\tilde{\mu}$ the field decays.  The maximum instability growth rate for
$a=0.99$ is $M\omega_I\approx3.6\times10^{-4}$ (with the growth rate for the
energy being twice that) and is achieved for $\tilde{\mu}\approx 0.47$.  For
sufficiently small $\tilde{\mu}$, we expect $M\omega_I \propto \tilde{\mu}^7$
while at larger $\mu$ the instability rate should switch signs (and become a
decay rate) when $\omega_R=m\Omega_H$.  Combining this with the above function
$\omega_{\rm Fit}(\mu)$, and fitting for the overall factor (which we find to
be close to unity) $A\approx1.0$ using the unstable points, we obtain a decent
approximation for the growth or decay rate of the dominant Proca mode in this
regime with  
\beq
\left[M\omega_I\right]_{\rm Fit} = A(\tilde{\mu}/m)^{4m+3}\left[a-2r_+\omega_{\rm Fit}/m\right],
\label{eqn:omegaI}
\eeq
where here $m=1$.
This is the dotted line in the bottom panel of Fig.~\ref{fig:omega_a99}.    

\subsubsection{Spin dependence}
\begin{figure}
\begin{center}
\includegraphics[width=\columnwidth,draft=false]{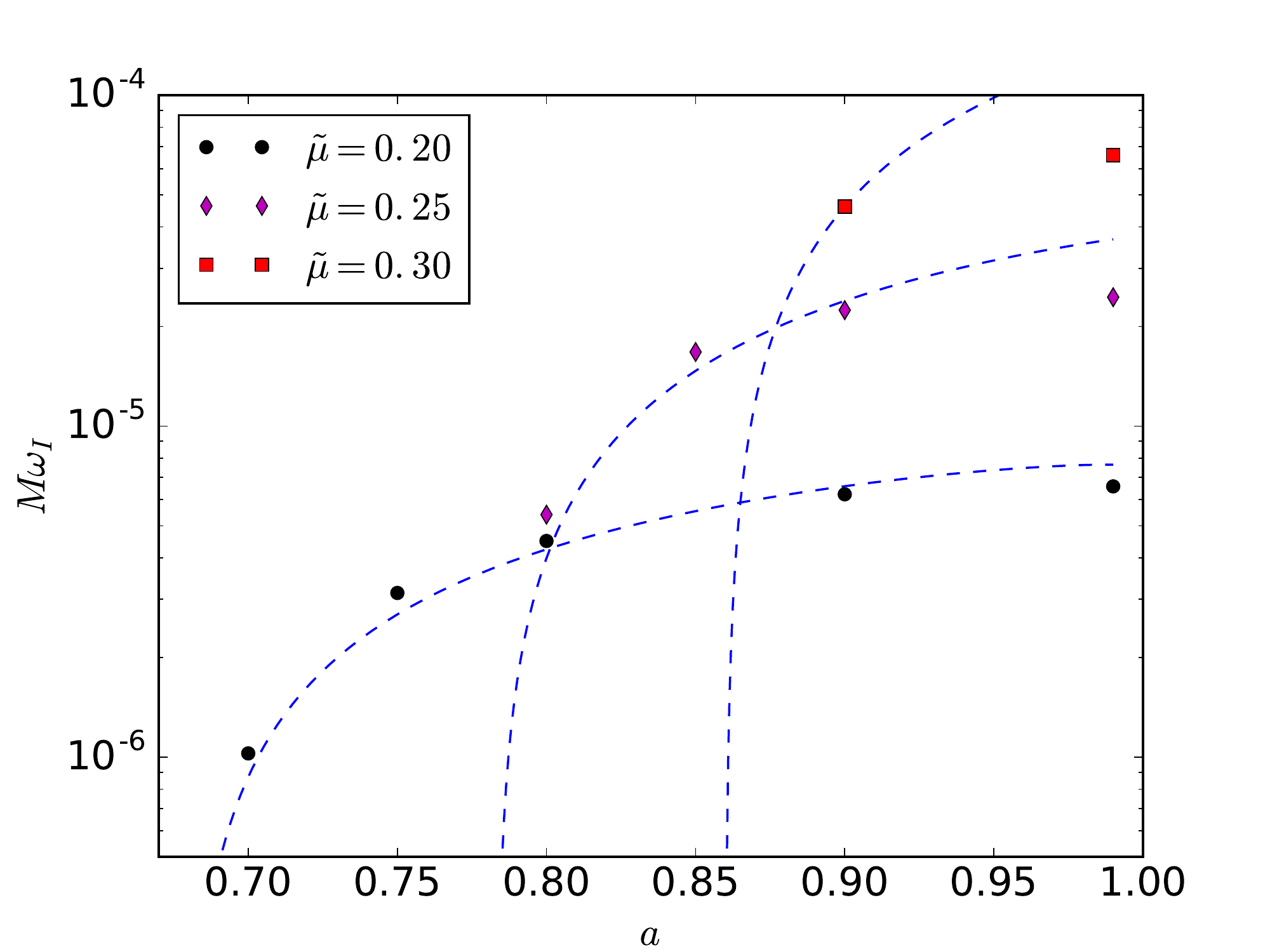}
\end{center}
\caption{
The $m=1$ instability growth rate as a function of black hole spin
for several values of the Proca mass.  The dotted blue
line shows the dependence $\omega_I \propto r_+(1-\omega_R/\Omega_H)$
with the amplitude fit to the $a\leq0.9$ points.
\label{fig:omega_spin}
}
\end{figure}

We also consider the behavior of the instability growth rate as a function of
black hole spin. Fixing $m=1$ and considering several values of $\tilde{\mu}$, we
obtain the results shown in Fig.~\ref{fig:omega_spin}.  For a fixed value of
$\tilde{\mu}$, the instability rate decreases with decreasing spin, and when
$\Omega_H<\omega_R/m$, superradiance shuts off.  However when $\omega_R$ is not
near $m\Omega_H$, the dependence of the instability rate on spin is not that
strong; for example, for $\tilde{\mu}=0.2$, the instability rate only decreases 
by $\sim50\%$ going from $a=0.99$ to $a=0.8$.  

\begin{figure}
\begin{center}
\includegraphics[width=\columnwidth,draft=false]{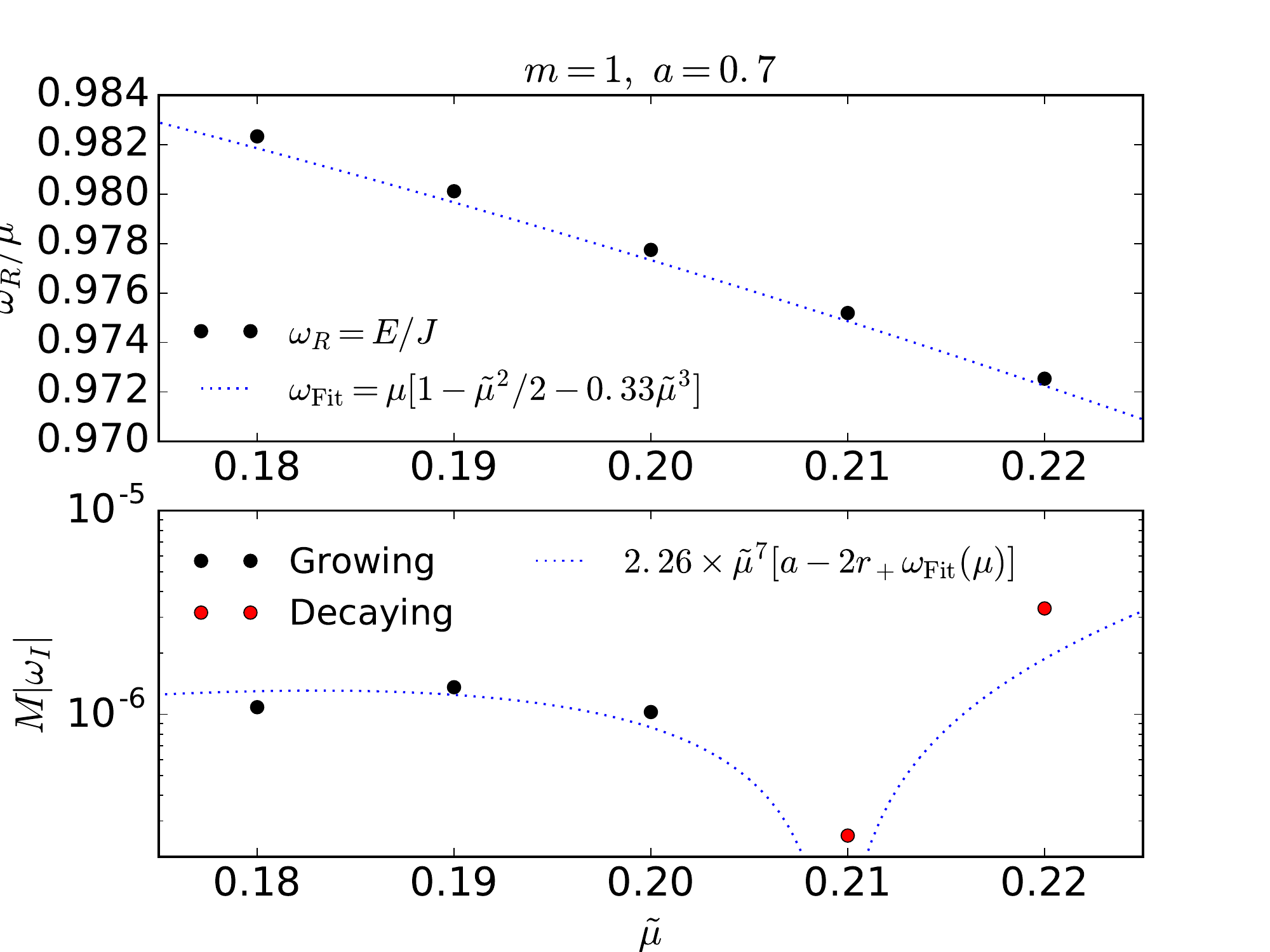}
\end{center}
\caption{
The same as Fig.~\ref{fig:omega_a99}, but for a black hole with spin $a=0.7$.
The dotted blue line in the top panel ($\omega_{\rm Fit}$) is the fit from the
$a=0.99$ data in Fig.~\ref{fig:omega_a99} (not the data shown in this figure).
\label{fig:omega_a7}
}
\end{figure}

For comparison in Fig.~\ref{fig:omega_spin} we also show an extrapolation of
the spin dependence of the instability of the form $\omega_I \propto
r_+(1-\omega_{\rm Fit}/\Omega_H)$. The choice of this functional form from
among other functions that decrease with decreasing $a$ (for
the values of $\mu$ shown) and go to zero when
$\omega_R\rightarrow \Omega_H$ was made more or less arbitrarily, and it seems
to overestimate the spin dependence at large spins for the higher Proca field
masses, but provides a decent match for $\tilde{\mu}=0.2$.  In
Fig.~\ref{fig:omega_a7} we show the real and imaginary frequency of the
dominant Proca field modes for $a=0.7$.  The dependence of frequency on $\mu$
is also approximately captured by Eq.~(\ref{eqn:omegafit}) obtained above from the
$a=0.99$ results, which is also shown in Fig.~\ref{fig:omega_a7}.   
For the instability rate, we also fit the functional form given by Eq.~(\ref{eqn:omegaI})
to these few unstable points and find a higher coefficient than in the $a=0.99$ case, 
$A\approx2.3$. 

\subsubsection{Modes with $m=2$}
\begin{figure}
\begin{center}
\includegraphics[width=\columnwidth,draft=false]{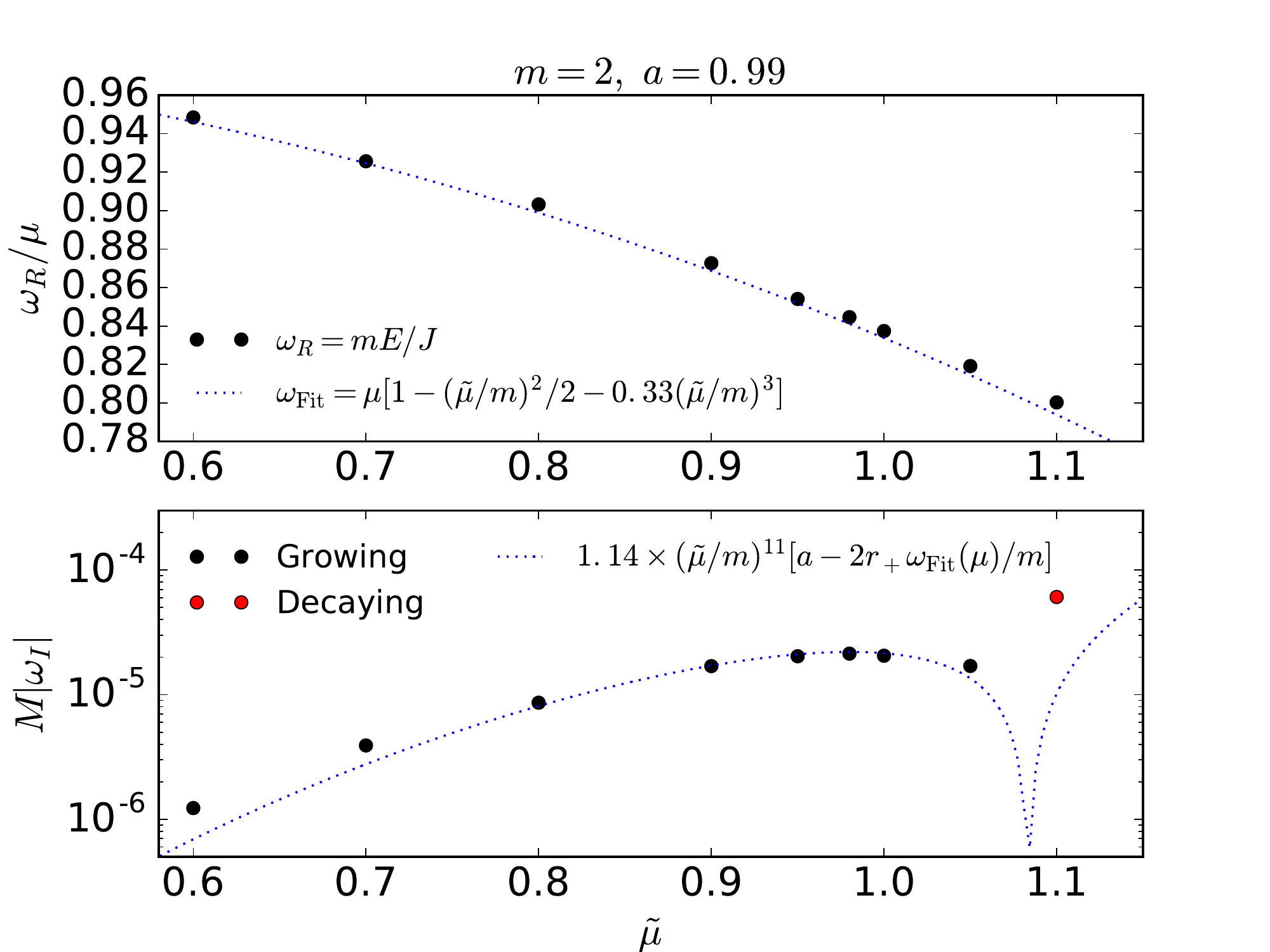}
\end{center}
\caption{
Similar to Fig.~\ref{fig:omega_a99}, but for modes with $m=2$ (and $a=0.99$).
Top panel: The real part of the frequency as measured from $\omega_R=2E/J$,
along with Eq.~(\ref{eqn:omegafit}) (which was obtained by fitting to the data in Fig.~\ref{fig:omega_a99}).
Bottom panel: The growth (black points) or decay (red points) rate of the dominant mode of the Proca field
along with the fitting function. 
\label{fig:omega_m2}
}
\end{figure}

Next we consider Proca field modes with one higher azimuthal number:
$m=2$.  The maximum instability rate for these modes is significantly smaller
than for $m=1$, though they are unstable for larger values of $\tilde{\mu}$.
We show results for $a=0.99$ and various values of $\tilde{\mu}$ in
Fig.~\ref{fig:omega_m2}.  Here we find that Proca fields with
$\tilde{\mu}\lesssim1.05$ are unstable. The maximum growth rate for $m=2$ and
$a=0.99$ is $M\omega_I\approx2\times10^{-5}$ and is achieved for $\tilde{\mu}\approx 0.98$. 
This is roughly a factor of $20$
slower than the $m=1$ case. Again we find that $\omega_R$ is well
approximated by the 
expression in Eq.~(\ref{eqn:omegafit}), but now with $m=2$. 
In Fig.~\ref{fig:omega_m2} we also show the
approximation given by Eq.~(\ref{eqn:omegaI}) for the instability rate, this time with $m=2$,
which approximates the trend with a best-fit amplitude of $A\approx1.1$. 

\subsection{Unstable mode}
\begin{figure}
\begin{center}
\includegraphics[width=\columnwidth,draft=false]{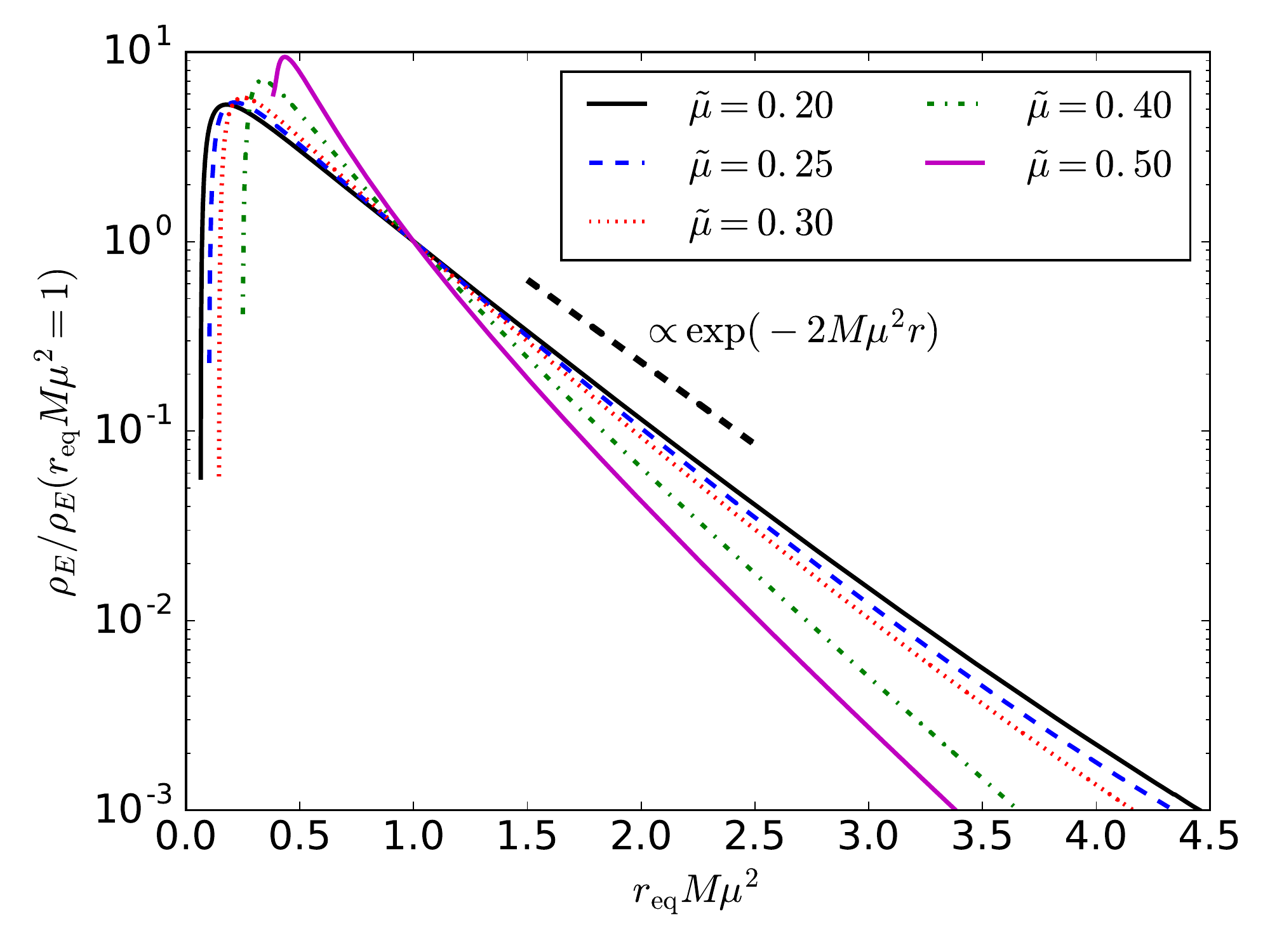}
\end{center}
\caption{
The falloff of energy density along the equator for cases with $m=1$, $a=0.99$ and
various values of $\mu$. The equatorial radius is multiplied by $M\mu^2$ which sets
the length scale, and for smaller values of $\mu$ the falloff of energy goes like $\exp(-2M\mu^2 r_{eq})$. 
\label{fig:e_falloff}
}
\end{figure}

In addition to the frequency and instability rate, we can also characterize the
spatial structure of the dominant unstable modes.  In general, for smaller
values of $\tilde{\mu}$, the modes will have larger characteristic length scales, and be
less concentrated near the black hole horizon.  This is illustrated in
Fig.~\ref{fig:e_falloff} where we show the energy density $\rho_E$ on the
equatorial plane for various values of $\tilde{\mu}$.  We can see that the Proca field
in the clouds falls off exponentially at large distances, and with
characteristic length $1/(M\mu^2)$ (or twice this value for the energy) for
smaller values of $\tilde{\mu}$, as expected for bound states in the hydrogen atom
approximation.  Here and below, the overall normalization of the unstable mode
is arbitrary.

\begin{figure}
\begin{center}
\includegraphics[width=\columnwidth,draft=false]{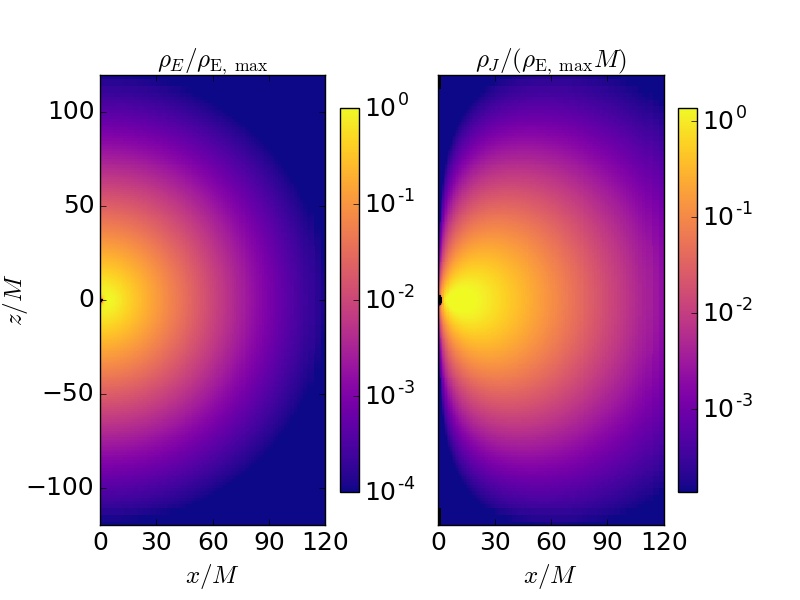}
\includegraphics[width=\columnwidth,draft=false]{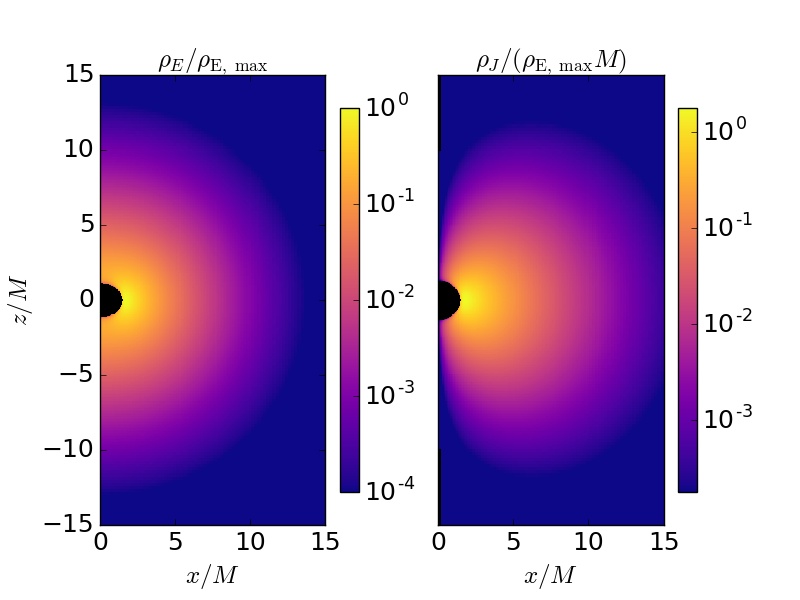}
\includegraphics[width=\columnwidth,draft=false]{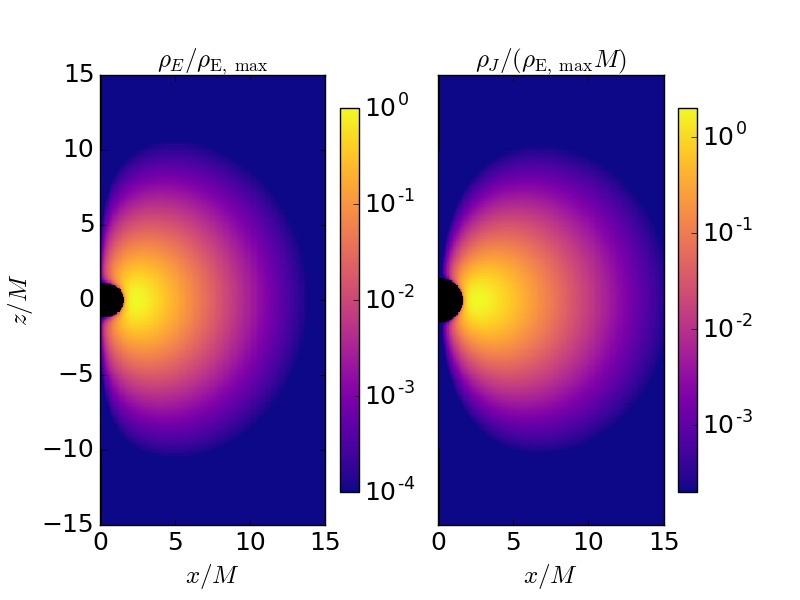}
\end{center}
\caption{
Energy (left) and angular momentum density (right) for cases with a black hole with $a=0.99$ and
(from top to bottom) $\tilde{\mu}=0.2$, $0.5$, and 0.95, and
$m=1$, 1, and 2, respectively.
The $z$ axis is the spin axis of the black hole and $z=0$ is the equatorial plane. 
Note the difference in spatial scales between the top and bottom two cases.
\label{fig:ej_color}
}
\end{figure}

\begin{figure}
\begin{center}
\includegraphics[width=\columnwidth,draft=false]{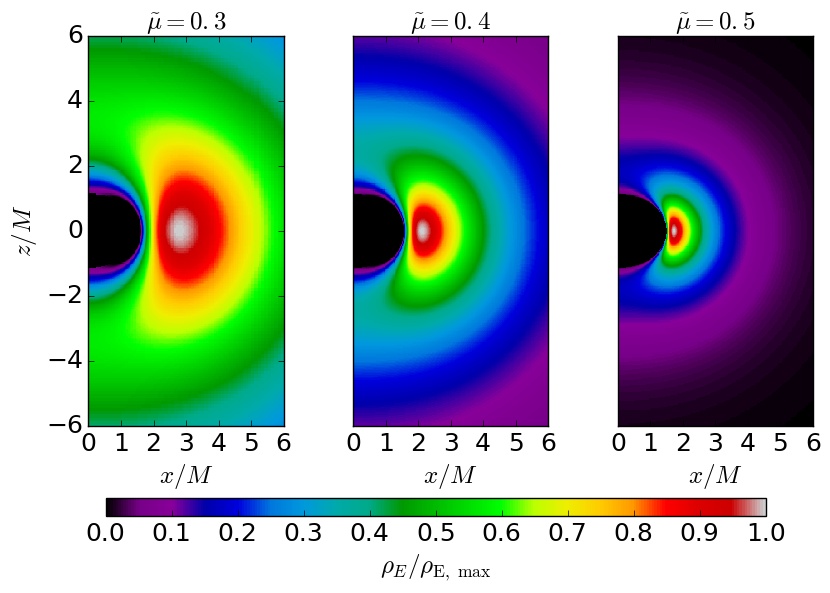}
\end{center}
\caption{
Energy density near the black hole for cases with $m=1$, $a=0.99$, and
$\tilde{\mu}=0.3$, 0.4, and 0.5 (left to right).
The $z$ axis is the black hole spin axis.
\label{fig:eclose_color}
}
\end{figure}

We also show the energy and angular momentum density for three representative
cases with $\tilde{\mu}=0.2$, $0.5$, and $0.95$, and $m=1$, 1, and 2, respectively, 
in Fig.~\ref{fig:ej_color}.  Farther
away from the black hole, and in particular for the lower $\mu$ cases, the
Proca cloud energy density is roughly spherical.  However, near the black hole the
Proca cloud shows deviations from sphericity.  This is illustrated in
Fig.~\ref{fig:eclose_color}, where it can be seen that as $\mu$ increases the
unstable mode becomes concentrated in a smaller and smaller region around the black hole
equator.  

\begin{figure*}
\begin{center}
\includegraphics[width=\columnwidth,draft=false]{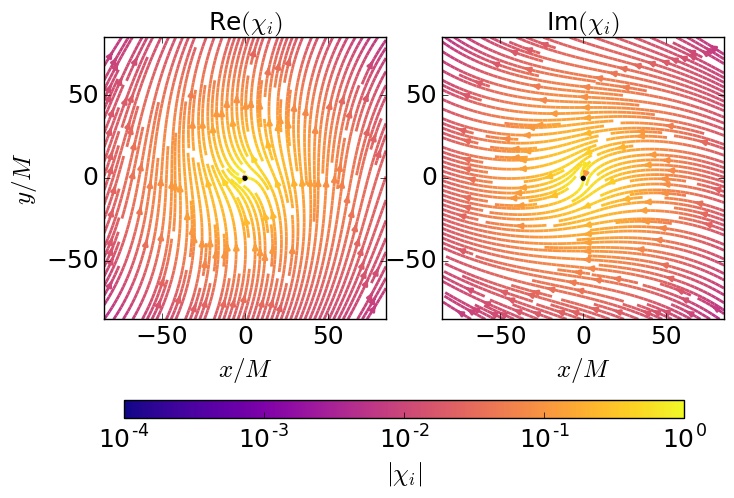}
\includegraphics[width=\columnwidth,draft=false]{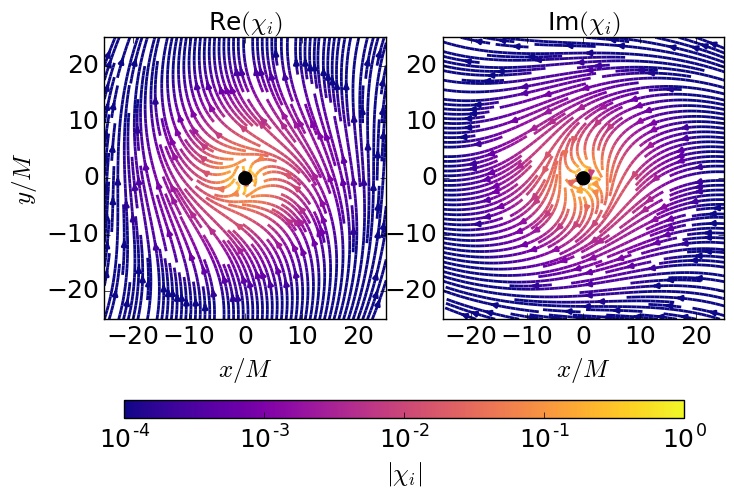}
\includegraphics[width=\columnwidth,draft=false]{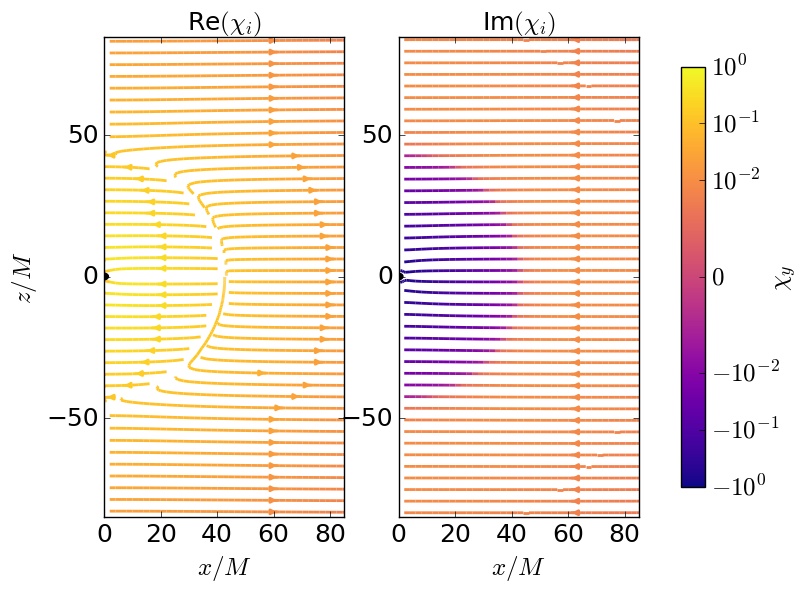}
\includegraphics[width=\columnwidth,draft=false]{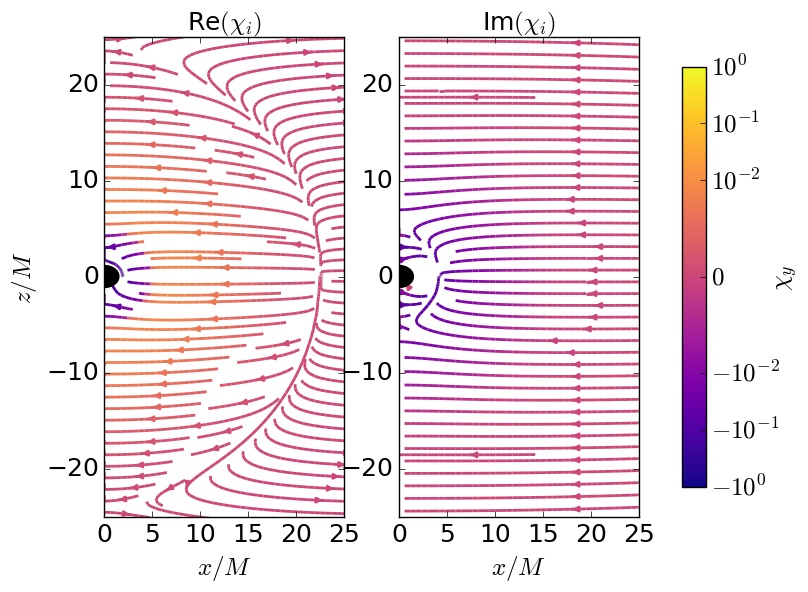}
\end{center}
\caption{
Streamlines of the real and imaginary parts of $\chi_i$ in the equatorial plane
($z=0$; top), and in the $y=0$ plane (bottom) for snapshots from $m=1$,
$a=0.99$, and $\tilde{\mu}=0.2$ (left) and 0.5 (right).
Note the difference in scales in the two cases.
\label{fig:chi_color}
}
\end{figure*}

We also show streamline plots for the spatial projection of the potential
vectors of the dominant unstable modes for representative cases in
Fig.~\ref{fig:chi_color}.  Over large distances, the vector field points in a
roughly constant direction (at a given instance of time), as expected from the
hydrogenic approximation.  In the vicinity of the black hole, the field lines
curve more strongly, though this may in part be an artifact of the particular
coordinates used here. 

\subsection{Gravitational waves}
\label{ssec:gw}
In this section we address what the GW signal would be
from a superradiantly unstable Proca field.  Above we have considered
configurations of a complex Proca field with an azimuthal symmetry which leads
to an axisymmetric stress energy tensor. 
At the test field level, this is mainly for computational convenience since
the real and imaginary parts of the Proca field are uncoupled. However, when 
coupled to gravity such a source will produce very little
gravitational radiation.  But if we instead consider just a single real
Proca field, the oscillation of the unstable mode around the black hole will produce
GWs at a frequency $2\omega_R$.  In order to calculate the magnitude of this,
we take the Proca field configuration from a simulation that has evolved
sufficiently long for the most unstable mode to dominate, and then use just the
real part of the Proca field as initial data for a fully three-dimensional
simulation where we evolve both the Proca field and the Einstein field
equations.  For all cases we scale the amplitude of the Proca fields so that the 
initial energy is $E/M\leq10^{-3}$, and ignore the self-gravity of the Proca field in the initial data
for the metric.  We evolve the Einstein equations using the background error
subtraction technique as described in~\cite{best} to minimize the truncation
error from just the spinning black hole solution.  After a brief transient period, this
produces a nearly monochromatic GW signal which we measure by extracting the
Newman-Penrose scalar $\Psi_4$.

\begin{figure}
\begin{center}
\includegraphics[width=\columnwidth,draft=false]{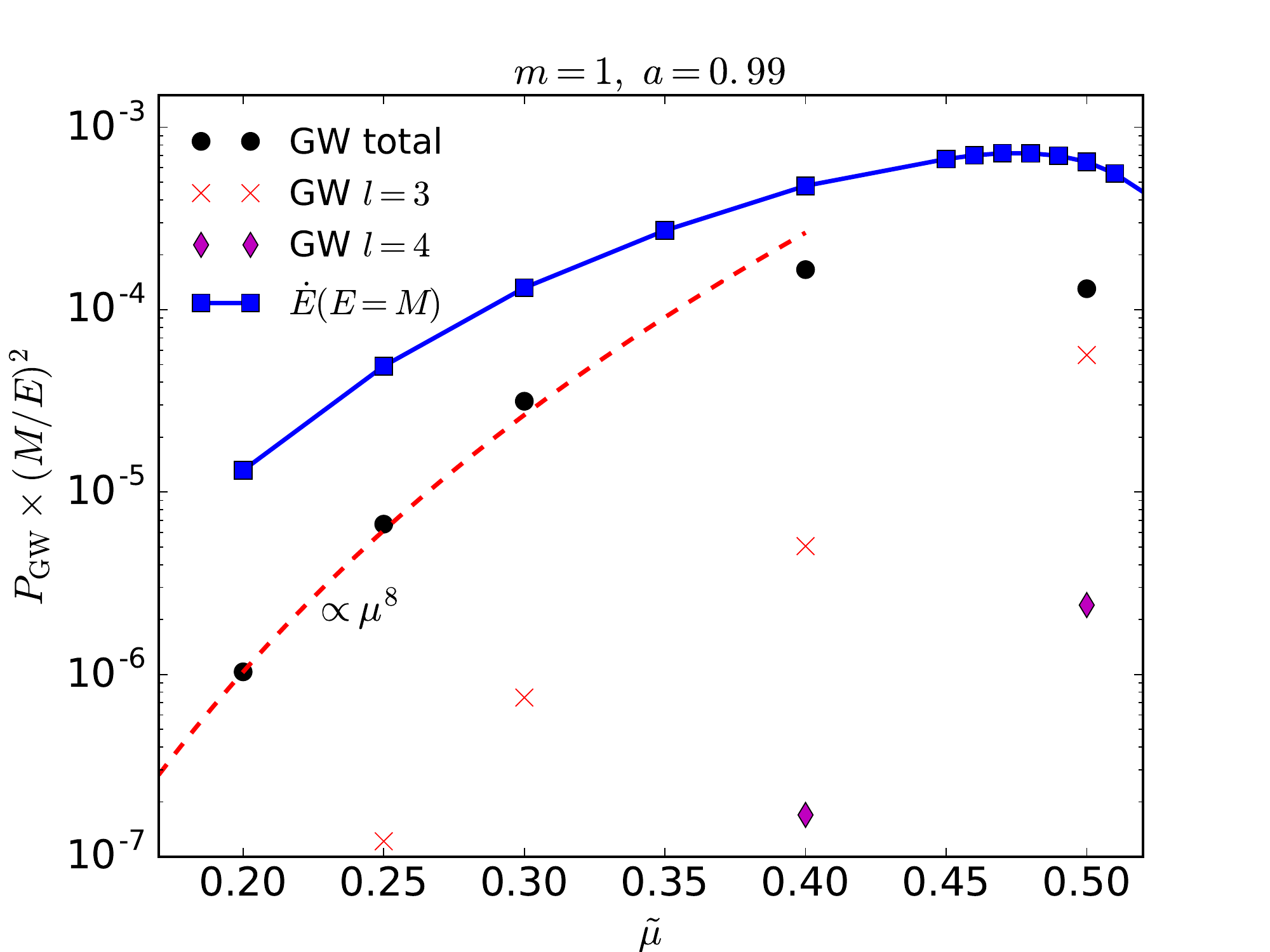}
\end{center}
\caption{
The GW power as a function of $\tilde{\mu}$ for the dominant $m=1$ unstable
mode consisting of a single real-valued Proca field around an $a=0.99$ black
hole (black points).  In addition to the total GW power, which predominately
comes from the $(\ell,m)=(2,\pm2)$ components,  we also show the contributions
from the $(\ell,m)=(3,\pm2)$ (red Xs) and $(\ell,m)=(4,\pm2)$ (magenta
diamonds) spin-2 spherical harmonics.  The GW power has been divided by the
square of the Proca field energy so as to be independent of the magnitude of
the mode, in the test field limit.  For comparison we also show the
superradiant instability growth rate of energy, at a reference energy of $E=M$
(blue squares).  The ratio of GW luminosity to the superradiant growth rate for
a Proca cloud of energy $E$ would then be given by multiplying the ratio of the
black points to the blue squares by $(E/M)$.
\label{fig:pgw}
}
\end{figure}

In Fig.~\ref{fig:pgw} we show the GW luminosity for $m=1$, $a=0.99$, and
several values of $\tilde{\mu}$
\footnote{
The resolution we use for these three-dimensional simulations to determine the GWs is equivalent
to the ``medium" resolution of~\cite{East:2013mfa}.  From one simulation at lower resolution for
the $\tilde{\mu}=0.4$, we estimate the truncation error in $P_{\rm GW}$ to be 
$<10\%$ and to primarily be an underestimate of the power due to the numerical dissipation 
of the GWs in the wave zone, which will be less severe for the lower $\tilde{\mu}$ cases, which have
longer wavelength GWs. 
}.  The GW luminosity has been scaled with the
square of the energy in the Proca field, so as to be independent of the Proca
field magnitude in the test field limit (which we have verified to
be approximately true for the values used here).  The gravitational radiation is
primarily quadrupolar, i.e. when performing a decomposition into spin-2 
spherical harmonics, the $\ell=|m|=2$ components dominate.  In
Fig.~\ref{fig:pgw} we also show the contribution from higher $\ell$ components
of the GW signal (still with $|m|$=2), which make a larger contribution for
larger $\mu$ cases where the Proca cloud is localized closer to the black hole
and less spherical.  For the highest mass parameter considered,
$\tilde{\mu}=0.5$, the $\ell=3$ contribution is comparable to the $\ell=2$.
This is similar to what was found for the dominant superradiantly unstable
mode in the massive scalar field case, though the overall magnitude of the GW
power found here for vector fields $P_{\rm GW}\sim10^{-4}(E/M)^2$ is
significantly larger than the $\sim10^{-8}$ values found in the scalar case for
$a=0.99$ and the regime of the maximum instability~\cite{Yoshino:2013ofa}.
For the smaller values of $\tilde{\mu}$ where the Proca cloud is concentrated
on scales large compared to the black hole radius, these results should not be very
sensitive to the black hole spin, and we have verified that for $\tilde{\mu}=0.2$, the
values of $P_{\rm GW}$ for $a=0.7$ and $a=0.99$ differ by $<10\%$.

Since the GW luminosity is proportional to the square of the Proca cloud
energy, while the superradiant growth rate is directly proportional to the
energy, the importance of GW emission will increase as the cloud grows.
However, as can be seen by comparison with the instability growth rate (also
shown in Fig.~\ref{fig:pgw} at a reference energy of $E=M$), it does not appear
that GW emission will significantly reduce the growth rate of the Proca cloud
due to superradiance, even up to the maximum saturation energies of
$E\approx0.09M$~\cite{nonlinear}, for these parameters.  Though we do not
consider very small values of $\tilde{\mu}$, the GW luminosity (divided by the
Proca energy squared) appears to be falling off more steeply with decreasing
$\mu$ than the instability rate---slightly steeper than $\tilde{\mu}^8$ from
the few points measured, versus $\tilde{\mu}^7$ for the instability rate---indicating that GW emission will be less significant at lower values of the
mass parameter.  The calculation of the GW power in the small $\tilde{\mu}$
limit performed in~\cite{BLT2017} indicates that $P_{\rm GW}\times(M/E)^2$
should fall off even more steeply, $\propto \tilde{\mu}^{10}$ in that regime.
The value for the GW luminosity we find here for $\tilde{\mu}=0.2$ falls
between the estimates obtained in the ``flat-space" approximation and the one
including ``Schwarzschild background corrections" in~\cite{BLT2017}.  

These results indicate that once the superradiant instability has extracted
enough rotational energy from the black hole for the horizon frequency to
decrease to the point where superradiance shuts off, the resulting Proca cloud
can dissipate through GW radiation, in many cases faster than the time scales on
which the higher $m$ superradiant instabilities will operate.
 
\section{Discussion and Conclusion}
\label{discussion}
We have used evolutions of the Proca equations on black holes spacetimes to
study the superradiant instability for a range of parameters, finding that the
growth rate for the field can be as high as $3.6\times10^{-4}M^{-1}$ and
$2\times10^{-5}M^{-1}$ for modes with $a=0.99$ and $m=1$ and 2  azimuthal
numbers, respectively.  For larger values of the Proca mass, near where the
instability growth rate is maximized, the unstable modes are concentrated in
the strong-field regime near the black hole horizon, and relativistic effects can be
important.  For the cases considered here (the fastest growing $m=1$ modes for
spins $0.7\leq a \leq 0.99$ and $m=2$ modes for $a=0.99$) we have found that
the frequency of the bound modes is well approximated by $\omega_R \simeq \mu
\left[1-(\tilde{\mu}/m)^2/2-(\tilde{\mu}/m)^3/3\right]$ where the last term is
an additional relativistic correction to the hydrogenic approximation that we
have rounded to a convenient fraction.  This relativistic effect extends the
range of mass parameters which are superradiantly unstable, and will also
determine the black hole horizon frequency at which the instability will
saturate~\cite{nonlinear} and be encoded in the frequency of any GWs sourced by
such modes.

This can be used to look for, or place constraints on, the existence of
ultralight vector fields using observations of spinning black holes, for
example from observations of x-ray binaries, or from observing GW signals.  As
an example of how the results obtained here might translate into astrophysical
terms, we note that a black hole formed from the merger of two equal mass,
nonspinning black holes has a spin of approximately $a\approx0.7$, which is
consistent, within a $90\%$ confidence interval, with the final black hole
spins measured in the first two LIGO detections GW150914 and
GW151226~\cite{Abbott:2016blz,Abbott:2016nmj}. Using the above results for
GW150914 (GW151226), the postmerger black hole with mass 62 $M_{\odot}$ (20.8
$M_{\odot}$) would be superradiantly unstable with $\tilde{\mu}=0.18$ if
there existed a massive vector boson with a physical mass of $4\times10^{-13}$
eV ($1\times10^{-12}$ eV).  It would take roughly 180 $e$-folds or 7 hours (2
hours) for a Proca cloud to grow from a single particle, to the level where it
had liberated the $E\approx 0.018M$ rotational energy from the black hole such
that the horizon frequency matches $\omega_R$.  The resulting Proca cloud would
emit GWs at a frequency $f_{\rm GW}=\omega_R/\pi\approx180$ Hz (550 Hz) with
power $P_{\rm GW}\sim10^{-10}\sim10^{50}$ erg/s for time scales of $E/P_{\rm
GW}\sim 11$ hours (4 hours).  On a much longer timescale of years, the $m=2$
superradiant instability could also grow, and spin the black hole down even
further.

This example is primarily for illustration, as the work of~\cite{BLT2017}---which explores the phenomenology of massive vector superradiance in 
detail---suggests that observations of x-ray binaries with rapidly spinning black holes
have already excluded these particular vector boson masses, and more massive
and/or rapidly spinning black hole mergers will be required to place further
constraints.

\subsection{Comparison to other works}
In this section we briefly compare the results obtained here to those in the
literature.  
The methods used here are very similar to those in~\cite{Witek:2012tr}, which
also used simulations to follow the evolution of Proca fields on a black hole
spacetime, except without assuming an azimuthal symmetry as we do here.
In~\cite{Witek:2012tr} superradiant instability was observed in one simulation
with $a=0.99$ and $\tilde{\mu}=0.4$ that exhibited strong beating of multiple
modes with a quoted instability rate of $M\omega_I\sim (5 \pm 1)\times10^{-4}$---or
half that value, depending on the quantity measured.  Here we
obtain that $M\omega_I\approx2.4\times10^{-4}$ for the same parameters, consistent with
the latter, and also find a value of $M\omega_R\approx0.36$ that matches the lower
frequency component found there.

\begin{figure}
\begin{center}
\includegraphics[width=\columnwidth,draft=false]{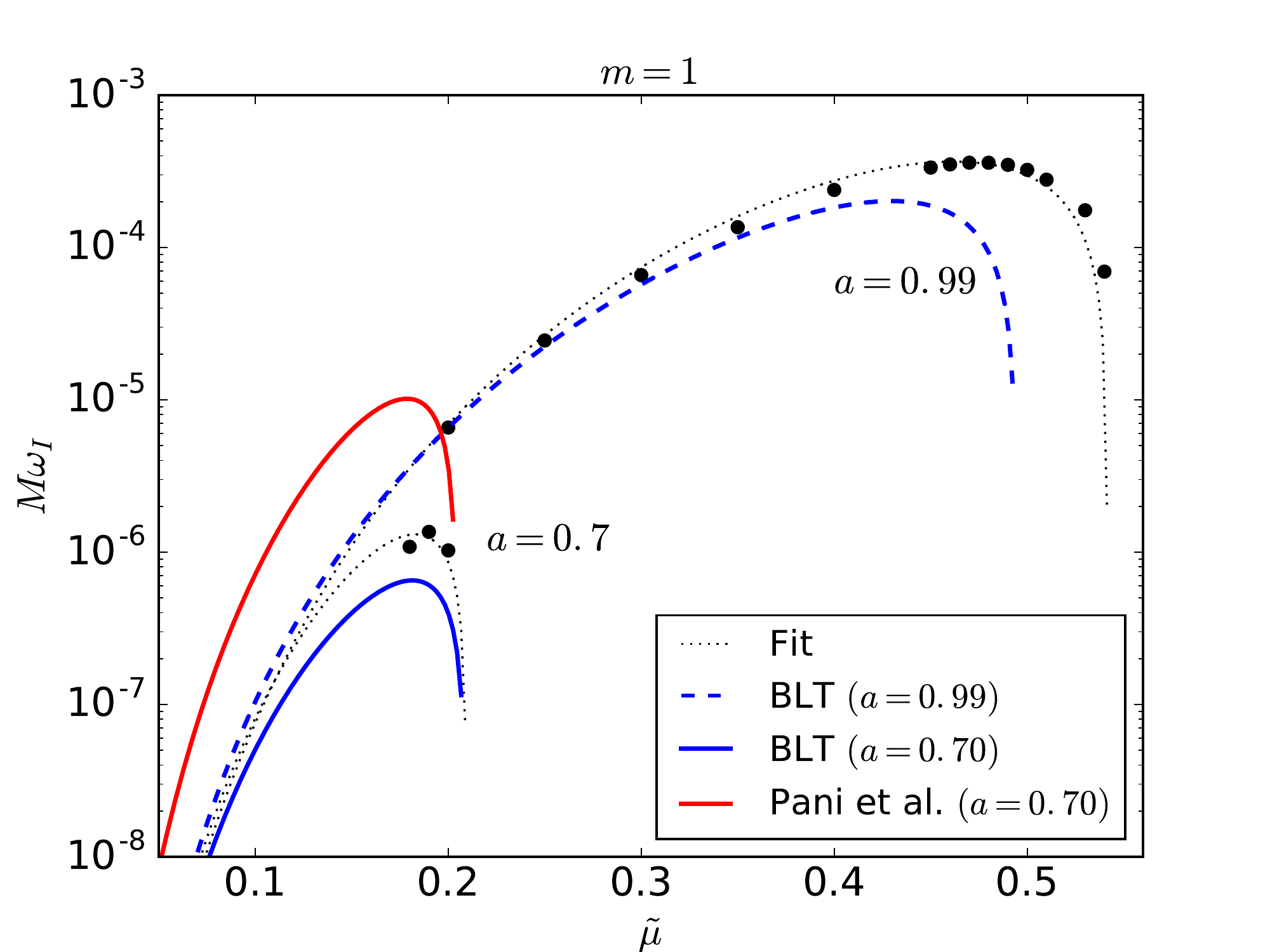}
\end{center}
\caption{
A comparison of the instability growth rate for the dominant $m=1$ mode
obtained here [black dots which are fit by the dotted curve, Eq.~(\ref{eqn:omegaI})] to 
the nonrelativistic approximation of~\cite{BLT2017} (blue curves)
and the low-spin approximation of~\cite{Pani:2012bp} (red curve) at two values of the black hole
spin $a=0.7$ and 0.99.
\label{fig:comp}
}
\end{figure}

In this paper we have focused on studying the superradiant instability in the
regime where the black hole spin and Proca mass parameter are not too small, since
the long instability time scales would otherwise be too computationally expensive to
follow with time-domain simulations.  Nevertheless, we can also compare our
results to those obtained in the nonrelativistic regime to see if there is a
region of overlap, as we do in Fig.~\ref{fig:comp} for the fastest growing
($m=1$) mode.  In~\cite{Pani:2012bp}, the authors study linear perturbations of
massive vector fields around slowly rotating black holes and fit the
instability rate they find to a function like the one used here [their Eq.~(98)]; however, the overall coefficient they find for small spins is roughly an
order of magnitude larger than the ones found here for large spins (which do increase with
decreasing spin).
Recently~\cite{BLT2017} have used a matching calculation to analytically
compute superradiance rates in the nonrelativistic limit which is valid
at small $\tilde{\mu}$ but arbitrary spin.  As shown in Fig.~\ref{fig:comp}, their
results agree quite well with those found here at $a=0.99$ and moderate values
of $\tilde{\mu}$, and the difference is subpercent level at $\tilde{\mu}=0.2$.
At high values of $\tilde{\mu}$, the simulation results give faster instability rates,
partly due to the relativistic corrections which reduce $\omega_R$. 
We also note that despite the good
agreement of the fitting function Eq.~(\ref{eqn:omegaI}) with the instability
rate of~\cite{BLT2017} in the regime of moderate $\tilde{\mu}$, the latter is
actually larger by a factor of $\approx2$ in the limit of
$\tilde{\mu}\rightarrow 0$, indicating that these fitting functions do not
fully capture the nonrelativistic limit.  For lower values of black hole spin, it does
appear that lower values $\tilde{\mu}$ have to be used to get similar levels of
agreement. For $a=0.7$ (also shown), we are not able to probe values
$\tilde{\mu}$ much below maximum instability, where the differences are more
pronounced (though the fit extrapolated to small $\tilde{\mu}$ is actually closer).  
In~\cite{Endlich:2016jgc} the instability rate is computed using
an effective field theory approach, also assuming small $\tilde{\mu}$, and a similar
scaling, but a larger coefficient is obtained.  See~\cite{BLT2017} for a more
detailed comparison of these different methods.

\subsection{Concluding remarks}
\label{conclusion}
We have explored the superradiant instability in the relativistic regime,
determining the growth rate, frequency, gravitational radiation and other
features of the dominant unstable modes. We have extended relativistic time-domain
simulations to sufficiently large Compton wavelengths where they overlap with
analytic results in the nonrelativistic limit, at least for rapidly spinning
black holes.  Though here we have focused on only the most unstable mode for the lowest
two azimuthal numbers $m$, techniques similar to those used
in~\cite{Dolan:2012yt} could be used to uncover a larger spectrum of modes from
time simulations like the ones performed here.  It would also be interesting
to see if the relatively simple form for the instability growth rate and
frequency found here could be derived analytically as a relativistic correction
in perturbation theory.  

In this study we have ignored the backreaction of the Proca field on the black
hole spacetime, except to study the gravitational radiation in the weak field
regime. This test field treatment will no longer be valid when the Proca grows
large enough to extract a non-negligible fraction of the black hole's rotational
energy. This regime is studied in~\cite{nonlinear} using these same methods as
here, but evolving the full Einstein-Proca equations for select cases with a
complex Proca field and an $m=1$ azimuthal symmetry.  It is found that the
superradiant instability smoothly saturates once the black hole has lost enough
angular momentum for its horizon frequency to match the frequency of the Proca
cloud that forms.\footnote{This is likely connected to the black hole with
Proca hair solutions found in~\cite{Herdeiro:2016tmi}; see
also~\cite{Sanchis-Gual:2015lje,Bosch:2016vcp} for studies of the analogous
charged black hole case.}  Another potential direction for future work is to
study higher azimuthal number instabilities, as well as the possibility for 
interactions between different unstable modes making up a Proca cloud, in the
nonlinear regime. 

For the case of a single real-valued Proca field, we have found that the
massive vector field hair that develops around a spinning black hole can efficiently
radiate away through GW emission in the relativistic regime, on time scales that
are only a few orders of magnitude longer than it took to develop.  It would
also be interesting to see how the details of the growth of the Proca cloud
and decay through gravitational radiation play out in the fully nonlinear case.

\acknowledgments
I thank Asimina Arvanitaki, Sam Dolan, Stephen Green, Luis Lehner, Jonah
Miller, and Frans Pretorius for stimulating discussions.  I am grateful to
Masha Baryakhtar, Robert Lasenby, and Mae Teo for sharing their related work as
it progressed.  Simulations were run on the Perseus Cluster at Princeton
University and the Sherlock Cluster at Stanford University.  This research was
supported in part by Perimeter Institute for Theoretical Physics. Research at
Perimeter Institute is supported by the Government of Canada through the
Department of Innovation, Science and Economic Development Canada and by the
Province of Ontario through the Ministry of Research, Innovation and Science.  

\appendix
\section*{Appendix: Convergence results}

\begin{figure}
\begin{center}
\includegraphics[width=\columnwidth,draft=false]{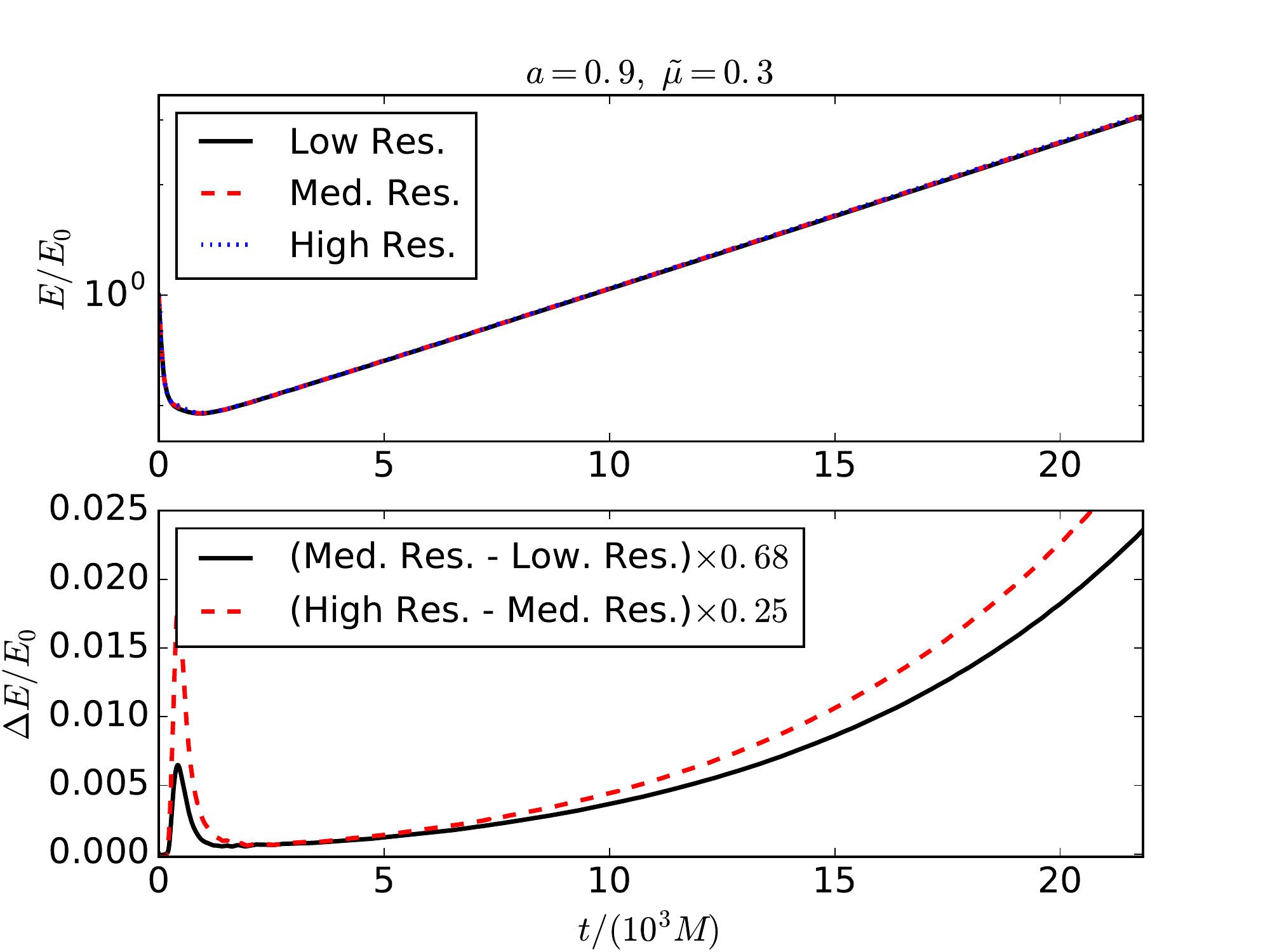}
\end{center}
\caption{
Top: The energy in the Proca field as a function of time for 
a case with $\tilde{\mu}=0.3$, $m=1$ and a black hole with spin $a=0.9$, at three different resolutions.
Bottom: The difference in this quantity between resolutions, scaled assuming fourth-order convergence.
The feature at early times is just truncation error from the numerical integration of the 
Proca energy, zeroed inside the black hole horizon, while a sizable fraction of the initial perturbation
is falling into the black hole.
\label{fig:rich}
}
\end{figure}

To establish numerical convergence and estimate truncation error, we run select
cases at three different resolutions, where the lowest has $dx\approx0.0256M$
on the finest grid, and the medium and high resolutions have $3/4$ and
$0.5\times$ smaller grid spacing, respectively.  For all cases we use seven levels of
mesh refinement, with 2:1 refinement ratio, centered on the black hole.  In
Fig.~\ref{fig:rich}, we show the Proca field energy as a function of time for an
example case at these three different resolutions. As expected, the error in
this quantity is consistent with between third- and fourth-order convergence
(with the quantities in the figure scaled assuming the latter). 

As an estimate of the truncation error in computing the instability growth
rate, we list the value of $\omega_{I}$ found by fitting the energy to an
exponential curve for several different cases, both as computed with the
lowest resolution, and the Richardson extrapolated value from all three
resolutions.  The difference between the two values gives the approximate truncation
error in this quantity at the lowest resolution.   We can see that the error is
greater for larger values of the mass parameter and spin. In the results
presented in this paper, we use at least the medium resolution when
$\tilde{\mu}/m\geq0.4$ (for $m=1$ and 2), and so we also list the values
obtained at this resolution in Table~\ref{tab:table1} for higher mass
parameters.  With these choices for resolution, the truncation error in
measuring $\omega_I$ is $\lesssim 5\%$ for the cases listed.  The Proca
field angular momentum shows very similar behavior to the energy, and in fact
the ratio of these quantities at late times varies very little with resolution;
e.g.  for the cases mentioned in this section, $E/J$ for the low and high
resolutions differ by $<0.01\%$.

\begin{table}[h!]
  \centering
  \caption{A comparison of the instability growth rate for various cases at low (and medium for select cases) resolutions,
          to the Richardson extrapolated value using all three resolutions.  This gives a measure of the truncation error.}
  \label{tab:table1}
  \begin{tabular}{c|c|c|c|c}
    \hline
    $\tilde{\mu}$ & $m$ & $a$ & $M\omega_I$: Low res. (med res.) & Rich. ext.\\
    \hline
    0.3 & 1 & 0.90 & $4.59\times10^{-5}$ & $4.60\times10^{-5}$\\
    0.3 & 1 & 0.99 & $6.58\times10^{-5}$ & $6.62\times10^{-5}$\\
    0.4 & 1 & 0.99 & $2.36\times10^{-4}$ & $2.31\times10^{-4}$\\
    0.5 & 1 & 0.99 & $3.0\times10^{-4}$ ($3.2\times10^{-4}$) & $3.3\times10^{-4}$\\ 
    0.95 & 2 & 0.99 & $1.6\times10^{-5}$ ($1.9\times10^{-5}$)& $2.0\times10^{-5}$\\ 
  \end{tabular}
\end{table}

In addition to truncation error, there will also be systematic error due to the
presence of other modes in addition to the most unstable one, confounding the
measurement. In general, this will bias the measured growth rate towards lower
values, and the frequency towards higher values.  This is more difficult to estimate
since it depends on how much the initial perturbation excites the other modes,
and what their growth is relative to the most unstable mode. The effects of
this will decrease the longer each case is involved, but will likely be the
dominant source of error for the cases with slower growth $\omega_I \lesssim
10^{-5}M^{-1}$, since we do not consider evolutions longer than $\sim 10^{5}M$
for computational reasons.  As a rough indication of this, we note that for the
smallest $\tilde{\mu}$ (leftmost) points in Figs.~\ref{fig:omega_a7}
and~\ref{fig:omega_m2}, if we instead use half the simulation time to measure
the instability rate, we obtain values that are $0.01\%$ and $4\%$ lower,
respectively. 

\bibliographystyle{aipauth4-1}
\bibliography{ref}
\end{document}